\documentclass[12pt,preprint]{aastex}

\shorttitle{Origin of The Sun and Planets}
\shortauthors{Yao}

\begin{document}

\title{A New Hypothesis On The Origin and Formation of The Solar And Extrasolar Planetary Systems}

\author{Lihong Yao}
\affil{Department of Physics, University of Virginia, Charlottesville, VA 22904, U.S.A.}
\email{lihongyaoastro@gmail.com}

\affil{Submitted to Icarus. Comments are welcome.}

\begin{abstract}

A new theoretical hypothesis on the origin and formation of the solar and extrasolar planetary systems is summarized and briefly discussed in the light of recent detections of extrasolar planets, and studies of shock wave interaction with molecular clouds, as well as H. Alfven's work on Sun's magnetic field and its effect on the formation of the solar system (1962). We propose that all objects in a planetary system originate from a small group of dense fragments in a giant molecular cloud (GMC). The mechanism of one or more shock waves, which propagate through the protoplanetary disk during the star formation is necessary to trigger rapid cascade fragmentation of dense clumps which in turn collapse quickly, simultaneously, and individually to form multi-planet and multi-satellite systems. Magnetic spin resonance may be the cause of the rotational directions of newly formed planets to couple and align in the strong magnetic field of a younger star. 
\end{abstract}


\section{Summary of The New Hypothesis} \label{newhypo}

\begin{figure}
\epsscale{0.7}
\plotone{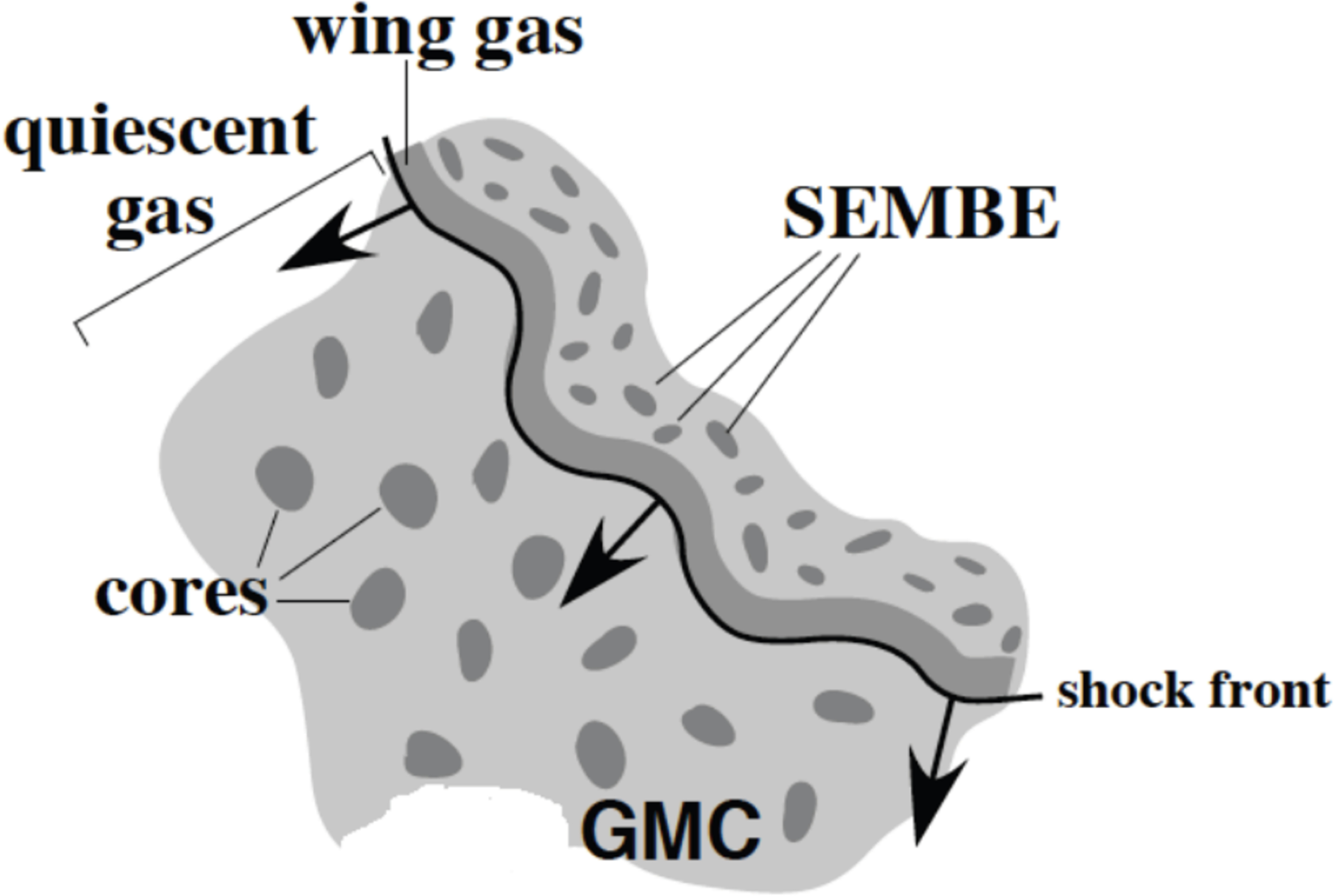}
\caption[]
{Schematic view of W44 shock wave propagates through the GMC and dense clumps detected by HCO+ J = 1 $\rightarrow$ 0 and CO J = 3 $\rightarrow$ 2 observations. Figure is obtained from Sashida et al. (2013), see details in Figure 8 of their paper. 
\label{sngmc}}
\end{figure}

\begin{figure}
\epsscale{0.7}
\plotone{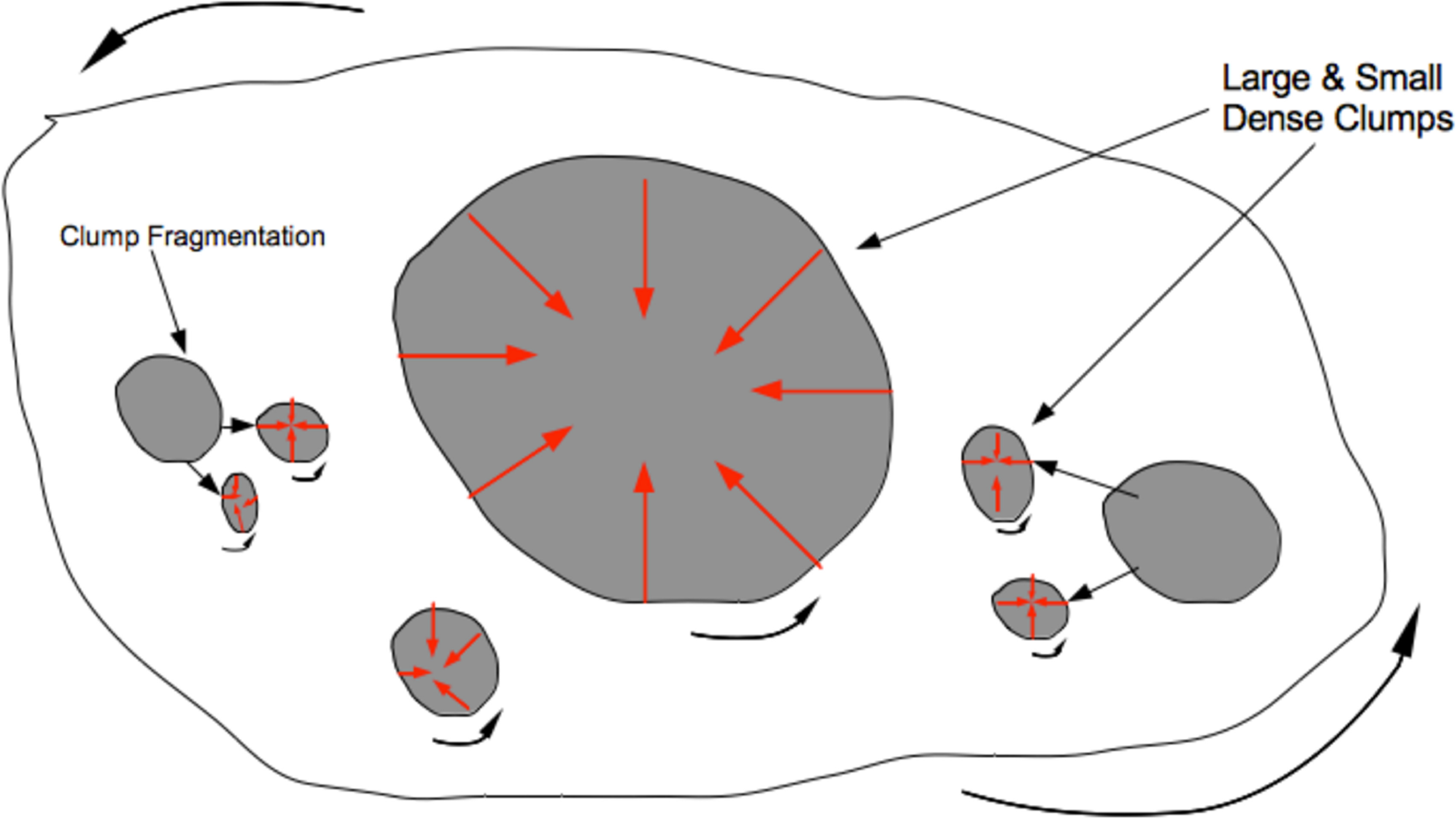}
\caption[]
{Schematic of rotation of a small group of dense clumps in a GMC; clump fragmentation and rotation of protostellar and protoplanet clumps.
\label{gmcsc}}
\end{figure}

The lifetime of giant molecular clouds and massive stars may be short, but the impact of their birth and death on the surrounding intercloud (ICM) and interstellar mediums (ISM) is profound \citep[and reference therein]{yao09, zwa10}. The strong stellar winds and supernova (SN) explosions from hundreds to thousands of the massive stars create a rapid expanding hot bubble. The kinetic energy in such supersonic expansion is thermalized by a stand-off shock. The high pressure downstream drives a strong shock into the ambient ISM, or even breaks through the cloud into the adjacent ICM. These strong and fast shock waves contribute significantly to the formation and rotation of the GMCs. On the other hand, thermal and dynamical instabilities produce strong cooling and turbulence in shocked gas on small scales, which play a dominant role in providing a non-linear density perturbation through the cloud \citep[e.g.][]{hei08}. This mechanism is needed during a GMC formation to create supersonic turbulence and to allow rapid fragmentation and formation of dense clumps locally and simultaneously before the onset of global gravitational collapse of the cloud. As observed in recent studies, for example, the interaction between W44 supernova remnant (SNR) and the adjacent GMC \citep[e.g.][and references therein]{cne95, fie11, sas13}, a secondary shock wave may produce further cascade fragmentations, hence formation of smaller and higher density clumps in the cloud (see Fig.~\ref{sngmc} and Fig.~\ref{gmcsc}). These small dense cores in turn quickly collapse to form stars or a star cluster \citep{shu87, smi09}. When a cloud (or a clump) has an initial non-zero angular momentum it will collapse to an accretion disk surrounding its central mass. Observations show that the mass spectrum of the star clusters follows the same power law distribution as that of the GMCs in the ISM \citep[e.g.][]{sol97, san99, ket05}. 

Core accretion and disk instability are the two currently accepted theories explaining how solar planets form \citep[e.g.][and reference therein]{hay85, mat07, hel13}. Core accretion model requires a few to a hundred million years to form a gas giant planet through the accretion of planetesimals from the protoplanetary disk, depending on the mass of the planet and how far it is to the parent star. This is a major problem for core accretion theory, because the time it takes to form a super giant planet (greater than a few times of the Jupiter mass) is much longer than the observed disk dissipation timescale of younger stars. For example, how can extrasolar planet GJ 504b, which has four times the Jupiter mass and about 160 million years old, form through planetesimals' collision and then core accretion at an orbital distance of 43 AU to its parent star GJ 504 \citep{kuz13}? The core accretion model also presents a problem for various planet migration theories \citep[e.g.][and reference therein]{anm11}. On the other hand, disk instability can form a super giant planet in a few hundred years, but the theory assumes that the protoplanetary disk is massive enough and can cool quickly enough to allow fragmentation to occur \citep{bos13}. In addition, sources and mechanisms for rapid fragmentation on small scales is needed to overcome global gravitational instability in the disk \citep{hei08}. Neither core accretion nor disk instability can explain the spin directions of solar planets, the conservation of angular momentum of the protostars, the evaporation of the protodisk, the formation of gas giant planet, and incredibly stable near-circular planetary orbits for billions of years, as well as why some low mass stars have planets while others have nothing around them, not even dust rings.

In this study, we propose that 4.6 billion years ago our Sun along with its planets and satellites were born from a small group of dense clumps, located in one of the star clusters with thousands of stars \citep[and references therein]{cam62, elm06, zwa09}. Like many other group of dense clumps, this group of clumps was initially compressed by a strong shock wave. The densest and largest clump quickly collapsed and formed a protostar of the Sun. Other compressed clumps that orbit around the protostar will eventually collapse due to gravitational instability (i.e. disk instability theory). However, if there is a secondary shock wave, e.g. an expanding thin shell of cold gas (produced by a nearby star formation) propagates into the disk before the onset of global gravitational collapse of dense clumps in the protoplanetary disk, the dense clumps behind the shock will then be further compressed. These compressed clumps may undergo rapid cascade fragmentation which break into smaller ones, and in turn become the seeds of multiple planets and their satellites.

\begin{figure}
\epsscale{0.7}
\plotone{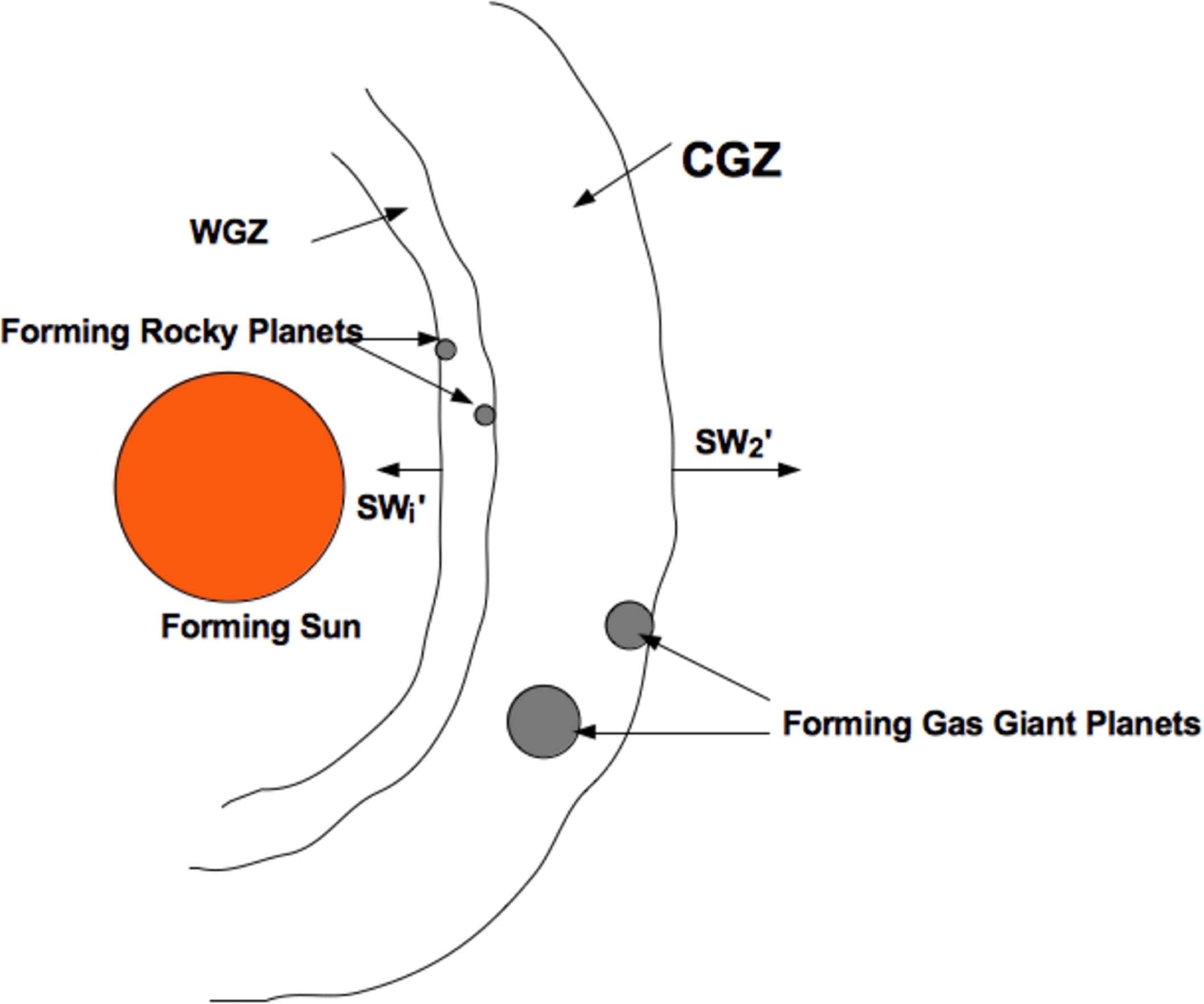}
\caption[]
{Schematic of expanding shocked gas regions produced by two reflected shock waves (SW$_i$$^\prime$ and SW$_2$$^\prime$) around a forming star. Rocky planets form in warm gas zone (WGZ), while gas giant planets form in cold gas zone (CGZ). 
\label{shockgas}}
\end{figure}

The secondary shock wave (SW$_2$) may be weaker than the supersonic shock wave that creates the GMC and star clusters, and it should have arrived at the solar protoplanetary disk before the Sun is completely formed. This shock wave sweeps up most of the interclump gas in the protoplanetary disk into its shell before it collides with the ionized gas shock (SW$_i$) front near the radius of $R_i$ = 0.277 AU corresponding to a 80 km s$^{-1}$ of critical velocity of ionized gas particles around the forming Sun \citep{anw62, alf62}. The denser fragments behind the shock waves quickly collapse and form protoplanets and protosatellites that orbit around them, as a result of strong thermal and dynamical instabilities that dominate on the small scales \citep{hei08, bnk13}. The shock waves are partially refracted and partially reflected. Hence, part of the cold gas in the shell of SW$_2$ merges with the warm and (partially) ionized gas of SW$_i$ leaving behind the reflected shock waves an expanding non-uniform refracted gas region (0.3 AU $\rightarrow$ 10 AU), where the cooling effect converts more warm gas into the cold gas (see Fig.~\ref{shockgas}). The reflected shock waves of SW$_2$ and SW$_i$ are rarefaction waves indicated by SW$_2$$^\prime$ and SW$_i$$^\prime$ in Fig.~\ref{shockgas}. The inward expansion of the refracted gas is stalled by the high pressure plasma gas front at R$_i$, so is the reflected shock wave of SW$_i$. The outward expansion that is confined by the reflected shock wave of SW$_2$  becomes the gas supply for building gas giant planets. The dense clumps in warm gas zone (WGZ, 0.3 AU - 2.7 AU) collapse instantly and form the rocky planets, i.e. Mercury, Venus, Earth, and Mars, while collapsing clumps in neutral cold gas zone (CGZ, 2.7 AU - 10 AU) accrete sufficient cold gas to form gas giant planets Jupiter, Saturn, and their satellites. During the shock collision of SW$_2$ and SW$_i$  very small clumps (smaller than the Earth moon) that are supposed to form the satellites of rocky planets may attain or lose their orbital velocities and escape. Some of these clumps may be thrown inward toward the star where they will be heated and ionized, while other clumps may be ripped into very tiny fragments and pushed outward into the discontinuous gas region (at $\sim$ 2.7 AU) between WGZ and CGZ. Eventually these tiny fragments from WGZ along with the condensed tiny cold gas clumps form the ring of the Asteroid belt (2 AU and 4 AU). The reflected shock wave of SW$_2$ continues propagating outward and beyond Uranus, Neptune, and Pluto, then gradually slows down and eventually stalls. It soon becomes fragmented and dispersed to be part of the cold icy gas rings (i.e. Kuiper belt), when it reaches the outer region of the solar system (30 - 50 AU from the Sun). This explains why materials from the early formation of the solar system are found in the Kuiper belt, when examining stardust samples return from Comet Wild 2 \citep[and reference therein]{das11}. 

In addition, during the stellar core accretion, much of the angular momentum is transferred from the forming star to its surrounding ionized gas, and then from the ionized gas to the angular momentum of protoplanets and their satellites during the head-on shock wave collision. The orbital parameters (especially their semi-major axis) for all planets and their satellites have  changed very little since the time of their formation.

\begin{figure}
\epsscale{0.9}
\plotone{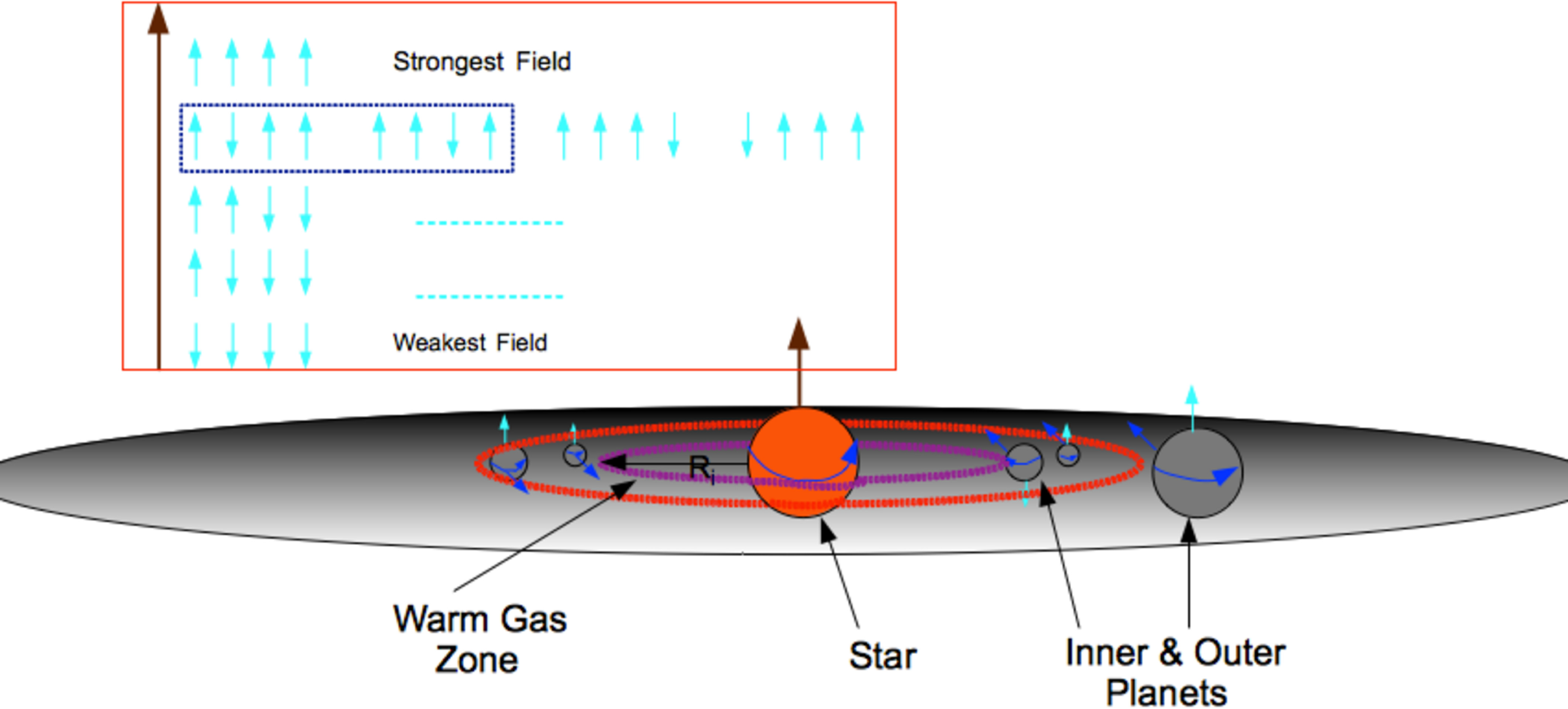}
\caption[]
{Schematic of spin-spin coupling of planets in an external strong magnetic field of a younger star. The blue-dashed box shows the spin-spin coupling of eight planets in solar system. 
\label{pspin}}
\end{figure}

A majority of the planets and their satellites have the same spin direction as the their orbital direction and spin direction of the Sun. If these planets and their satellites, the asteroids, were formed from merely random core accretions (i.e. planetesimal accretion theory), we would see an even mixture of directions of orbital revolution and rotation of planets and their satellites. However, if there is a certain mechanism as proposed in this study, which allows a small group of dense clumps that are orbiting around the forming Sun to undergo rapid fragmentation followed by a quick collapsing both locally and simultaneously, we are expected to see the formation of multiple planets and satellites like our solar system. Hence, the newly formed planets may experience a strong spin-spin coupling and begin to align each other in the external magnetic field generated by the plasma gas surrounding the forming parent star (see Fig.~\ref{pspin}). The spin axes of the planets will precess around the magnetic field. Magnetic spin resonance is the origin of the observed rotational directions of solar planets which follows the Pascal's rule of a quintet configuration (1:4:6:4:1), as indicated by blue-dashed box in Fig.~\ref{pspin}. This also implies that water may exist in all newly formed planets in our solar system. Magnetic field of planets may also play a dominate role in spin-spin coupling of large satellites at the time of their formation. The magnetic fields of the Sun and its planets may have evolved over the past 4.6 billion years, but the spin coupling among the planets remains strong and stable. On the other hand, orbital coupling or resonance in our solar system is governed by the gravitational influences between the parent star, the planets, and their satellites \citep[e.g.][]{gnp67, gre77}.

The initial conditions that determine the formation and final configuration of a star/planet system may vary from one group of dense clumps to another. Building such a stabilized system will strongly depend not only on the physical and chemical properties of its parent GMC and distribution of groups of dense clumps, but also on the triggering mechanisms for rapid fragmentation, core accretion, planet/satellite formation in the protoplanetary disk around a forming star. Therefore, each star/planet system may have its own unique characteristics depending on the initial conditions of the cloud, the clumps, and the ambient environment at the time of their formation. For example, recent detection of single planet systems found that HD 106906b (11 M$_{Jupiter}$) are located at a very large distance 650 AU to its parent star HD 106906 (K-type star) with only 13 million years age \citep{bai13}, and PSO J318.5-22 is a free-floating gas giant (6.5 M$_{Jupiter}$) formed without a parent star 12 million years ago \citep{mic13}. These two exoplanets may have formed directly from clump collapsing due to the onset of global gravitational instability in a cloud before they become fragmented. On the other hand, the HD 10180 is a multi-planet system; all of its planets are located at very close distances (0.02 - 3.5 AU) to their parent star HD 10180 (G1V star, 7.3 Gyr), but none of its planets has mass like gas giants (Jupiter or Saturn) \citep{tuo12}. The Gliese 876 system has four planets, two of them have Jupiter mass but are much closer to their parent star (0.1 - 0.2 AU) than those in our solar system \citep[and reference therein]{sha08}. The Gliese 876 system also has a notable orbital arrangement between its four planets, i.e. Laplace resonance configuration, which is similar to that of Jupiter's closest Galilean moons. On the other hand, four detected planets in the Upsilon Andromedae system are gas giant planets (0.6 - 4 M$_{Jupiter}$), located within 5 AU distance to their parent star \citep{hag08}. The star of the Upsilon Andromedae system is about 3.5 times more luminous than the Sun, and the distance of its inner most planet to its star is about six times closer than that of the Mercury to the Sun. Another multi-planet system that is similar to the Upsilon Andromedae system is the HR 8799 W system, but the inner most planet locates at a distance of 14 AU to its parent star that is 35 times further than that of the Mercury to the Sun \citep{opp13}. These extrasolar systems that have multiple planets may form in a similar way like our solar system, but have different clump properties, distribution, and protoplanetary disk conditions.

From what has been said here, a hypothesis is an idea. More theoretical computation and observational studies are needed to draw more definite conclusions. These studies may include the properties of small groups of dense clumps and the interclump conditions in the GMCs, the role of shock waves, the interplay among thermal, dynamic and gravitational instabilities on large and small scales, their effects on planetary system formation and evolution, as well as observational measurements of gas and dust properties in nearby protoplanetary disks around forming/younger stars, e.g. the HD 142527 system \citep{fuk13}.

\acknowledgments

I would like to thank my colleagues Dr. Claus Leitherer at Space Telescope Science Institute and Dr. Joe Ganem at Loyola University Maryland for their encouragement.

\end{document}